\documentclass[prb,aps, twocolumn, amsmath,amssymb,superscriptaddress]{revtex4}
\usepackage[dvips]{epsfig}
\usepackage{float}
\usepackage{amsmath}
\usepackage{amssymb}
\usepackage{amsfonts}
\usepackage{euscript}
\usepackage{enumerate}
\usepackage{hhline}
\usepackage{threeparttable}
\usepackage{multirow}
\usepackage{supertabular}
\usepackage{pslatex}
\usepackage{tabularx}
\usepackage{color}
\usepackage{graphicx}% Include figure files
\usepackage{dcolumn}% Align table columns on decimal point
\usepackage{bm}% bold math
\usepackage[sort&compress]{natbib}

\makeatletter
\renewcommand{\@biblabel}[1]{#1. }
\renewcommand{\@dotsep}{500}
\renewcommand{\@pnumwidth}{0em}
\renewcommand{\l@figure}[2]{% #1 is e.g. Figure 1 + caption, #2 is pg.
        \@dottedtocline{1}{1.5em}{2em}{Figure #1}{}\vspace{15pt}}

\begin{document}

\preprint{Draft 4}

\title{Optomechanical transduction of an integrated silicon cantilever probe using a microdisk resonator}% Force line breaks with \\

\author{Kartik Srinivasan} \email{kartik.srinivasan@nist.gov}
\affiliation{Center for Nanoscale Science and Technology, National
Institute of Standards and Technology, Gaithersburg, MD 20899, USA}
\author{Houxun Miao}
\affiliation{Center for Nanoscale Science and Technology, National
Institute of Standards and Technology, Gaithersburg, MD 20899, USA}
\affiliation{Maryland Nanocenter, University of Maryland, College
Park, MD 20742}
\author{Matthew T. Rakher}
\affiliation{Center for Nanoscale Science and Technology, National
Institute of Standards and Technology, Gaithersburg, MD 20899, USA}
\author{Marcelo Davan\c co}
\affiliation{Center for Nanoscale Science and Technology, National
Institute of Standards and Technology, Gaithersburg, MD 20899, USA}
\affiliation{Maryland Nanocenter, University of Maryland, College
Park, MD 20742}
\author{Vladimir Aksyuk} \email{vladimir.aksyuk@nist.gov}
\affiliation{Center for Nanoscale Science and Technology, National
Institute of Standards and Technology, Gaithersburg, MD 20899, USA}
\affiliation{Department of Electrical and Computer Engineering,
University of Maryland, College Park, MD 20742}

\date{\today}% It is always \today, today,

\begin{abstract}
\textbf{Sensitive transduction of the motion of a microscale
cantilever is central to many applications in mass, force, magnetic
resonance, and displacement sensing. Reducing cantilever size to
nanoscale dimensions can improve the bandwidth and sensitivity of
techniques like atomic force microscopy, but current optical
transduction methods suffer when the cantilever is small compared to
the achievable spot size.  Here, we demonstrate sensitive optical
transduction in a monolithic cavity-optomechanical system in which a
sub-picogram silicon cantilever with a sharp probe tip is separated
from a microdisk optical resonator by a nanoscale gap. High quality
factor ($Q\approx10^5$) microdisk optical modes transduce the
cantilever's MHz frequency thermally-driven vibrations with a
displacement sensitivity of $\approx4.4{\times}10^{-16}$
m/$\sqrt{\text{Hz}}$ and bandwidth $>1$ GHz, and a dynamic range
$>10^6$ is estimated for a 1 s measurement. Optically-induced
stiffening due to the strong optomechanical interaction is observed,
and engineering of probe dynamics through cantilever design and
electrostatic actuation is illustrated.}
\end{abstract}

 \maketitle

\begin{figure}[t]
\centerline{\includegraphics[width=\linewidth]{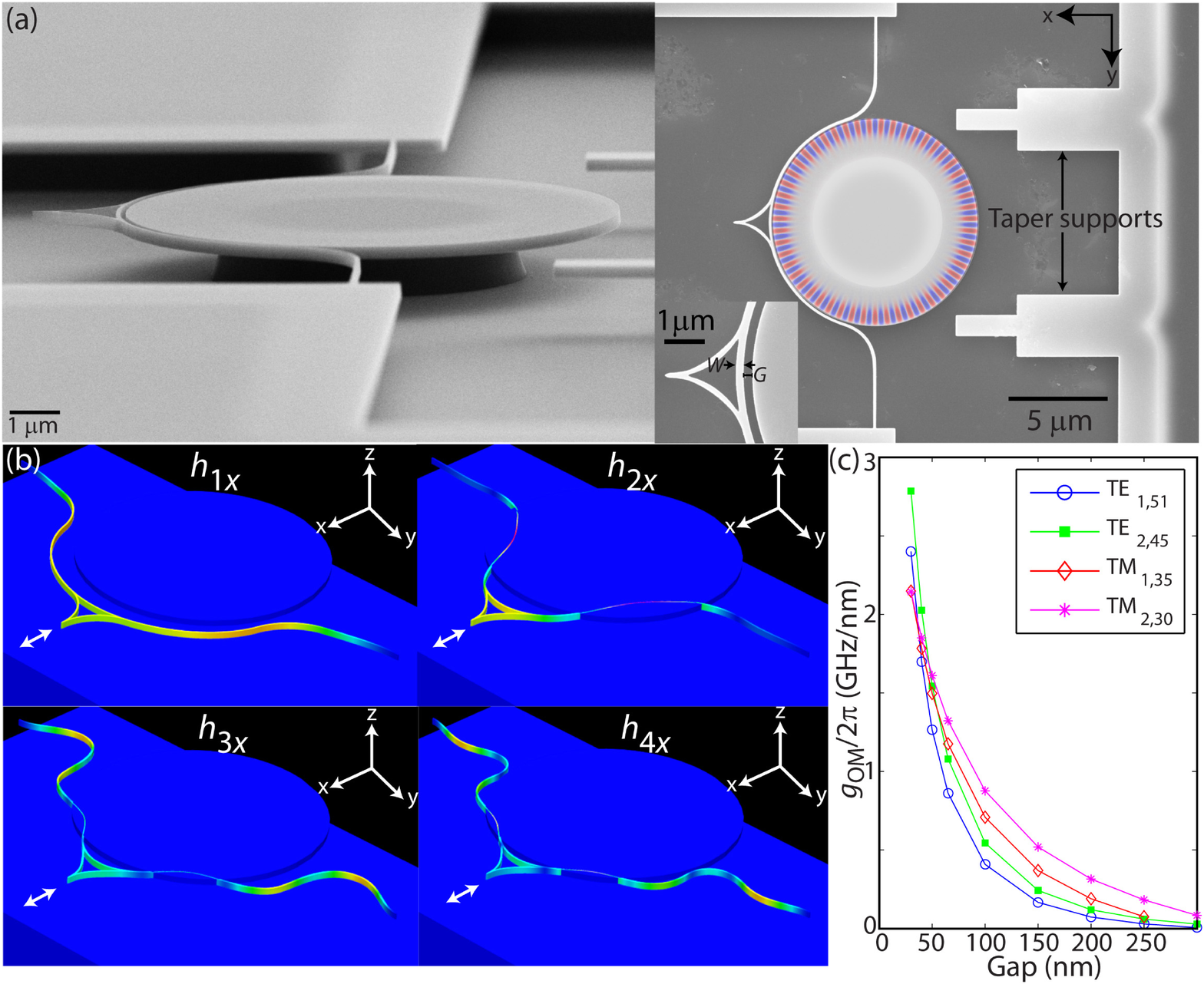}}
 \caption{(a) Scanning electron micrographs of the
 cantilever-microdisk system.  The right image has the FEM-calculated
 $z$-component of the magnetic field for the TE$_{1,51}$ mode overlaid on the structure,
 while the inset shows a zoomed-in region with cantilever width $W$
 and gap $G$; (b) Simulated mechanical modes (amplitude exaggerated for clarity)
with dominant displacement along the $x$-axis for $W$=65 nm; (c)
Predicted optomechanical coupling $g_{\text{OM}}$ between the
$h_{1x}$ cantilever mode and TE/TM modes of the microdisk.}
\label{fig:Figure1}
\end{figure}

\renewcommand{\arraystretch}{1.3}
\renewcommand{\extrarowheight}{0pt}
\begin{table*}
\caption{Calculated and measured properties of the $h_{mx}$
cantilever modes.} \label{table:cantilever_modes}
\begin{center}
\begin{tabularx}{0.73\linewidth}{lllllll}
\hline \hline
Mode & \quad $k$ (calc.) & \quad $m$ (calc.) & \quad $\Omega_{M}/2\pi$ (calc.) & \quad $\Omega_{M}/2\pi$ (expt.) & \quad $\Gamma_{M}/2\pi$ (expt.) & \quad $Q_{M}$ (expt.) \\
%\hhline{|=:=:|}
\hline
$h_{1x}$      & \quad 0.14 N/m  & \quad 0.73 pg & \quad 2.23 MHz  & \quad $2.35$ MHz         & \quad $479\pm8$ kHz & \quad 4.9 \\
$h_{2x}$      & \quad 1.41 N/m  & \quad 0.58 pg & \quad 7.82 MHz  & \quad $7.89$ MHz         & \quad $598\pm11$ kHz & \quad 13.1 \\
$h_{3x}$      & \quad 5.72 N/m  & \quad 0.35 pg & \quad 20.37 MHz & \quad $20.51$ MHz        & \quad $533\pm2$ kHz & \quad 38.5 \\
$h_{4x}$      & \quad 12.43 N/m & \quad 0.32 pg & \quad 31.17 MHz & \quad $31.36$ MHz        & \quad $706\pm4$ kHz & \quad 44.4 \\
$h_{5x}$      & \quad 41.95 N/m & \quad 0.44 pg & \quad 49.36 MHz & \quad $49.86$ MHz        & \quad $815\pm4$ kHz & \quad 61.2 \\
$h_{6x}$      & \quad 74.05 N/m & \quad 0.40 pg & \quad 68.13 MHz & \quad $68.71$ MHz        & \quad $752\pm43$ kHz & \quad 91.0 \\
\end{tabularx}
\end{center}
\end{table*}

Micro- and nanoscale cantilevers are at the heart of many
applications in mass, force, magnetic resonance, and displacement
sensing\cite{ref:Ekinci_Roukes_RSI,ref:Li_Tang_Roukes,ref:Craighead_mass_sensing,ref:Rugar}.
In atomic force microscopy (AFM)\cite{ref:Giessibl_RMP}, the push
towards smaller cantilevers\cite{ref:Walters_RSI,ref:Kawakatsu} is
motivated by the ability to increase mechanical frequencies while
maintaining a desired level of stiffness.  This influences the force
sensitivity and measurement bandwidth, in turn determining the image
acquisition rate and ability to resolve time-dependent forces and
acquire additional information about the tip-sample interaction
potential\cite{ref:Sahin}.  Standard optical methods for transducing
cantilever motion include beam deflection\cite{ref:Meyer_AFM} and
laser interferometry\cite{ref:Rugar_AFM}, and in macroscopic devices
that are 1 mm ${\times}$ 1 mm ${\times}$ 60 $\mu$m (length, width,
and height), quantum-limited displacement sensitivity of
$4{\times}10^{-19}$ m/$\sqrt{\text{Hz}}$ has been
achieved\cite{ref:Arcizet}. Interferometric approaches using a high
numerical aperture objective have also been used in micro-scale
devices, resulting in displacement sensitivities of
$3{\times}10^{-14}$ m/$\sqrt{\text{Hz}}$ for cantilevers that are
$20$ ${\mu}$m ${\times}$ 4 ${\mu}$m${\times}$ 0.2 ${\mu}$m and
$1{\times}10^{-15}$ m/$\sqrt{\text{Hz}}$ for larger conventional
cantilevers ($223$ ${\mu}$m ${\times}$ 31 ${\mu}$m${\times}$ 6.7
${\mu}$m)\cite{ref:Hoogenboom}. However, as the cantilever
dimensions are pushed below the detection wavelength, diffraction
effects limit the sensitivity of these
approaches\cite{ref:Ekinci_interferometer}, and near-field optics
and/or integrated on-chip detection methods can be of significant
benefit.

To that end, researchers have recently used evanescently coupled
on-chip waveguides\cite{ref:Povinelli} acting as doubly-clamped
cantilevers\cite{ref:Li_Tang2,ref:Roels} to demonstrate displacement
sensitivities of $3.5{\times}10^{-14}$ m/$\sqrt{\text{Hz}}$, while
end-to-end waveguides acting as singly-clamped
devices\cite{ref:Pruessner} have achieved similar
performance\cite{ref:Li_Tang_cantilever}.  Although these
waveguide-based approaches are optically broadband, the strong,
multi-pass interaction provided by optical cavities can be of
considerable advantage. Cavity
optomechanics\cite{ref:Kippenberg_Vahala_OE,ref:Favero_Karrai,ref:van_Thorhout}
has seen substantial recent progress, where in many cases the
optical resonator also acts as a mechanical oscillator, and its
internal vibrations have been transduced with measurement
imprecision at or below the standard quantum limit
\cite{ref:Schliesser_resolved_sideband2,ref:Teufel} and with
absolute displacement sensitivities in the $10^{-17}$
m/$\sqrt{\text{Hz}}$ to $10^{-18}$ m/$\sqrt{\text{Hz}}$
range\cite{ref:Eichenfield_zipper,ref:Schliesser_NJP}. In contrast,
here we focus on transducing the motion of a cantilever probe,
requiring a design in which the cantilever can be brought near a
surface and its fluctuations sensed by a nearby optical cavity
without inducing excessive optical loss.

A similar approach was presented in Ref.
\onlinecite{ref:Anetsberger_near_field}, where doubly-clamped
SiN$_{x}$ nanobeams were brought into the near-field of SiO$_2$
microtoroid cavities fabricated on a separate chip. In comparison,
here we fabricate a cantilever-optical cavity system on a single
silicon device layer, while tailoring the cantilever geometry for
both strong optomechanical interactions and applicability to AFM.
Beyond demonstrating sub-fm/$\sqrt{\text{Hz}}$ sensitivity to
cantilever motion, this approach has many potential benefits for
AFM. Silicon's high refractive index allows for significantly
smaller optical cavities to be used, yielding stronger
cantilever-cavity coupling rates and permitting higher bandwidth
operation. Moving to silicon opens up potentially advanced device
functionality, including electrostatic actuation and integrated
optical waveguide readout. By largely separating the mechanical and
optical designs, engineering of the cantilever geometry to achieve
desired parameters can be accomplished without adversely affecting
the optical readout mechanism. In addition, the strong
optomechanical interaction can allow for optical control of
cantilever mechanics, through effects such as optically-induced
stiffening and optically-driven mechanical
vibrations\cite{ref:Sheard,ref:Kippenberg_Vahala_OE,ref:Favero_Karrai,ref:Eichenfield_zipper}.
Finally, this platform provides simplifications with respect to
free-space detection systems that may improve measurement stability
and be of importance in parallelized multi-probe
measurements\cite{ref:Minne} or environments with limited optical
access.  This work lays the foundations for a class of practical
nanoscale mechanical sensors enabled by cavity optomechanics.

\noindent\textbf{Device geometry and simulation}

A simple device geometry is shown in Fig. \ref{fig:Figure1}(a), with
fabrication details given in the Methods. A semicircular cantilever
of width $W$ is suspended at its ends and separated by a gap $G$
from a 10 $\mu$m diameter silicon microdisk. The silicon is 260 nm
thick, and the cantilever has been designed to support a sharp tip
at its midpoint. Devices are fabricated with $W$=65 nm, 100 nm, and
200 nm, and nominal values $G$=50 nm, 75 nm, and 100 nm.  Scanning
electron microscope (SEM) images indicate that $W$ is typically
within $\pm$5 nm of its nominal value, while $G$ is often smaller
than the nominal value by a couple tens of nanometers, though
charging effects due to the electron beam limit this estimate. The
cantilever geometry is chosen to maximize its interaction with
microdisk optical modes while minimizing the scattering loss induced
by its presence. Optical modes are labeled TE$_{p,n}$ and
TM$_{p,n}$, according to polarization (transverse electric or
transverse magnetic) and radial ($p$) and azimuthal ($n$) order.
Three-dimensional finite element method (FEM) eigenfrequency
simulations indicate that, for $W$=65 nm or $W$=100 nm, cavity
quality factors ($Q$s) in excess of 10$^6$ can be achieved for
TE$_{1,n}$ and TE$_{2,n}$ modes, and $Q$s in excess of 10$^5$ can be
achieved for TM$_{1,n}$ modes, even as $G$ decreases to $\approx$30
nm. In comparison and as a baseline, fabricated microdisks without
cantilevers exhibit $Q$s in the mid-$10^5$ to low-$10^6$ range.

\begin{figure*}
\centerline{\includegraphics[width=0.8\linewidth]{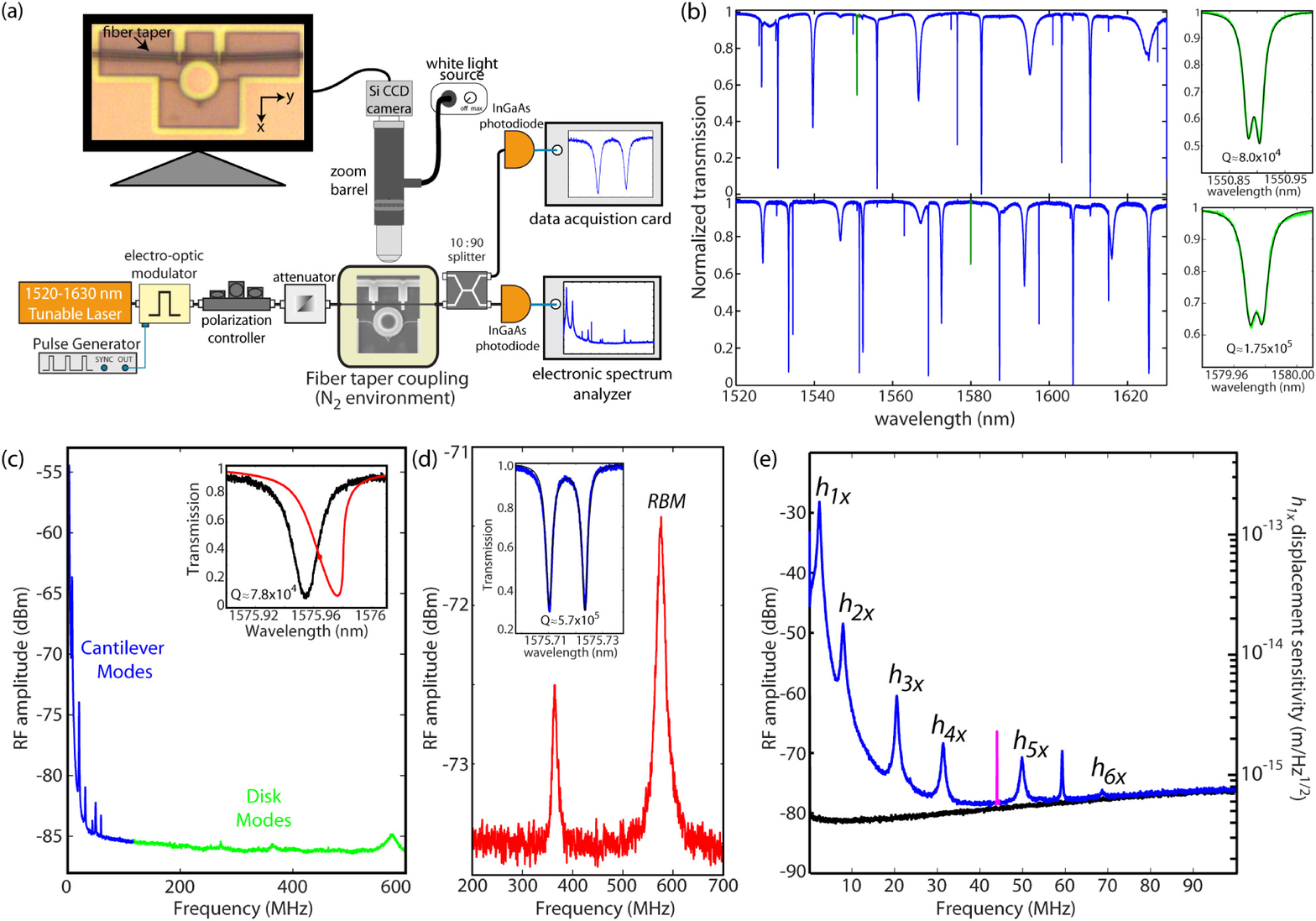}}
 \caption{(a) Setup for device characterization. (b) Broad wavelength
 scan (left) for TE (top) and TM (bottom) modes of a typical disk-cantilever
device ($W$=65 nm, $G$=50 nm).  Zoomed-in scans (right) show data
(green) along with a doublet model fit (black). (c) Broad RF
spectrum of a disk-cantilever device ($W$=65 nm, $G$=100 nm),
transduced by fixing the probe laser on the short wavelength side of
the TE-polarized mode shown in the inset (black=low power,
$P_{\text{in}}=14.1 \mu~W$, red=high power, $P_{\text{in}}=223
\mu~W$). Mechanical modes
 below 100 MHz (blue) are due to the cantilever, while modes
at $364.63$ MHz and $577.20$ MHz (green) are due to the disk. (d) RF
spectrum of a disk $\textit{without}$ the cantilever, displaying
modes at 364.74 MHz and 576.20 MHz. The inset shows a high-$Q$ TE
optical mode of the disk (blue) with fit (black). (e) Zoomed-in RF
spectrum of the disk-cantilever, showing the $h_{mx}$ modes (blue),
calibration peak (purple), and detection background (black).
\label{fig:Figure2}}
\end{figure*}

Mechanical modes of a $W$=65 nm cantilever are determined from FEM
simulations (see Methods), with representative modes shown in Fig.
\ref{fig:Figure1}(b).  We have focused on the $h_{mx}$ modes, which
are even symmetry in-plane modes whose primary displacement
direction is normal to the gap ($x$ direction), as they are the
dominant modes that are optically transduced and are of particular
relevance to AFM work. The predicted stiffness ($k$), resonant
frequency ($\Omega_{M}$), and effective mass ($m$) of these modes
are compiled in Table \ref{table:cantilever_modes}. Focusing on the
$h_{1x}$ mode at $\Omega_{M}/2\pi=2.23$ MHz, its optomechanical
coupling to $p$=1 and $p$=2 optical modes in the 1550 nm band,
defined as $g_{\text{OM}}=d\omega_{c}/{dG}$ ($\omega_{c}$ is the
cavity mode frequency), is calculated by FEM simulation and
displayed in Fig. \ref{fig:Figure1}(c). For the range of gaps
studied in this work, $g_{\text{OM}}/2\pi\approx0.5$ GHz/nm to
$g_{\text{OM}}/2\pi\approx3.0$ GHz/nm.  This is about two orders of
magnitude larger than $g_{\text{OM}}$ for SiN$_x$ cantilevers
coupled to SiO$_{2}$ microtoroids\cite{ref:Anetsberger_near_field},
and is due to the more tightly confined optical modes supported by
the silicon microdisks.

\noindent\textbf{Transduction of cantilever motion}

We measure the fabricated devices using a fiber taper coupling
method\cite{ref:Srinivasan7} shown schematically in Fig.
\ref{fig:Figure2}(a) (see Methods). A 1550 nm band tunable diode
laser is attenuated and coupled into the devices using an optical
fiber taper waveguide, a single mode optical fiber whose minimum
diameter has been adiabatically and symmetrically reduced to around
1 $\mu$m.  At this diameter, the waveguide mode's spatial profile
extends well beyond the glass core into the surrounding air
cladding, and this evanescent tail is used to excite and collect
light from the microdisk modes. The signal exiting the cavity is
split by a 90:10 fiber coupler, with 10 $\%$ of the light used for
monitoring the transmission level and recording swept-wavelength
transmission spectra, and 90 $\%$ sent into a radio frequency (RF)
photodetector, after which an electronic spectrum analyzer measures
RF oscillations in the detected signal.

Normalized transmission spectra over the full wavelength band for TE
and TM polarized modes of a $W=65$ nm, $G=50$ nm device are shown in
Fig. \ref{fig:Figure2}(b), along with zoomed-in scans of individual
modes. The polarization of the modes is determined by comparing the
free spectral ranges for modes of a given radial order with those
predicted from simulation. Loaded cavity $Q$s of $8.0{\times}10^4$
and $1.8{\times}10^5$ are observed for this device (corresponding
intrinsic $Q$s of $1.1{\times}10^5$ and $2.1{\times}10^5$,
respectively), which supports doublet modes due to
surface-roughness-induced backscattering that couples the clockwise
and counterclockwise modes of the cavity\cite{ref:Borselli2}.  Over
all devices, $Q$s of $5{\times}10^4$ to $2{\times}10^5$ are
typically observed for TE$_{1,n}$, TE$_{2,n}$ and TM$_{1,n}$ modes,
though occasional devices have $Q$s as high as
$\approx6{\times}10^5$ (see supplemental data). Optical transduction
of the cantilever's motion due to thermal noise is performed by
fixing the laser on the blue-detuned shoulder of a TE-polarized
cavity mode. A 1 MHz to 600 MHz spectrum for a $W=65$ nm, $G=100$ nm
device is shown in Fig. \ref{fig:Figure2}(c), and contains several
peaks. Those below 100 MHz originate from motion of the cantilever,
while those at higher frequencies ($364.63\pm0.35$ MHz and
$577.20\pm0.25$ MHz) are from motion of the disk. This is confirmed
by measuring the RF spectrum of a disk without a cantilever (Fig.
\ref{fig:Figure2}(d)) through a high-$Q$ cavity mode (loaded
$Q=5.7{\times}10^5\pm0.5{\times}10^5$, intrinsic
$Q\approx1.0{\times}10^6$), which yields RF peaks at near-identical
frequencies ($364.74\pm0.03$ MHz and $576.20\pm0.03$ MHz).  FEM
simulations indicate that the higher frequency mode is the disk's
radial breathing mode ($RBM$); its measured linewidth is
$\Gamma_{M}/2\pi=21.68\pm0.06$ MHz, corresponding to
$Q_{M}\approx27$.

Focusing on the frequency range between 100 kHz and 100 MHz, a
higher resolution RF spectrum at 223 $\mu$W of input power
($P_{\text{in}}$) into the cavity is shown in Fig.
\ref{fig:Figure2}(e). The frequencies of the transduced modes (Table
\ref{table:cantilever_modes}) correspond well with the previously
described simulation results. The mechanical quality factors of
these modes are between $Q_{M}\approx5$ for the $h_{1x}$ mode and
$Q_{M}\approx61$ for the $h_{5x}$ mode (Table
\ref{table:cantilever_modes}); these values are likely limited by
air damping\cite{ref:Verbridge}. The detection background, shown in
Fig. \ref{fig:Figure2}(e) in black, is found by placing the laser
off-resonance while maintaining a fixed detected power. Focusing on
the $h_{1x}$ mode, its calculated effective mass and measured
frequency correspond to a peak displacement amplitude of
$x_{rms}=\sqrt{k_{B}T/k}\approx160$ pm when driven by thermal noise
at 300 K. We use $x_{rms}$ and $\Gamma_{M}$ to convert the RF
amplitude in Fig. \ref{fig:Figure2}(e) to displacement
sensitivity\cite{ref:Ekinci_Roukes_RSI}. The corresponding
photodetector-limited sensitivity is
$4.4{\times}10^{-16}\pm{0.3\times}10^{-16}$ m$/\sqrt{\text{Hz}}$.
This value is consistent with that determined by a phase modulator
calibration (Methods) to within our uncertainty in the
disk-cantilever gap.  It represents an improvement by about a factor
of 100 with respect to other on-chip silicon cantilever
experiments\cite{ref:Li_Tang_cantilever,ref:Li_Tang2}, is at the
same absolutely sensitivity level demonstrated for SiN$_x$
cantilevers transduced by silica
microtoroids\cite{ref:Anetsberger_near_field}, and is about a factor
of 5 times larger than the standard quantum limit\cite{ref:Caves}
for our system. Along with the sensitivity, two other important
quantities that characterize this system for its use as a
displacement sensor are its dynamic range and bandwidth. The maximum
detectable displacement is approximately the ratio of the cavity
linewidth ($\Gamma/2\pi=2.44$ GHz) to $g_{\text{OM}}$, and is
$\approx4$ nm, giving a dynamic range $>10^6$ (60 dB) for a 1s
measurement. The bandwidth (BW) is limited by the cavity's response
time, which determines how quickly it can transduce mechanical
motion.  We therefore expect a BW$>1$ GHz, and this is substantiated
by transduction of the 575 MHz oscillations of the disk as
previously described in Fig. \ref{fig:Figure2}(c)-(d). Adjusting the
BW (e.g., through the waveguide coupling) allows for gain/BW
tradeoff within the fixed gain-BW product.  The large BW of these
devices is one advantage of relying on large $g_{\text{OM}}$ rather
than ultra-high-$Q$ for displacement detection.

\begin{figure}[t]
\centerline{\includegraphics[width=\linewidth]{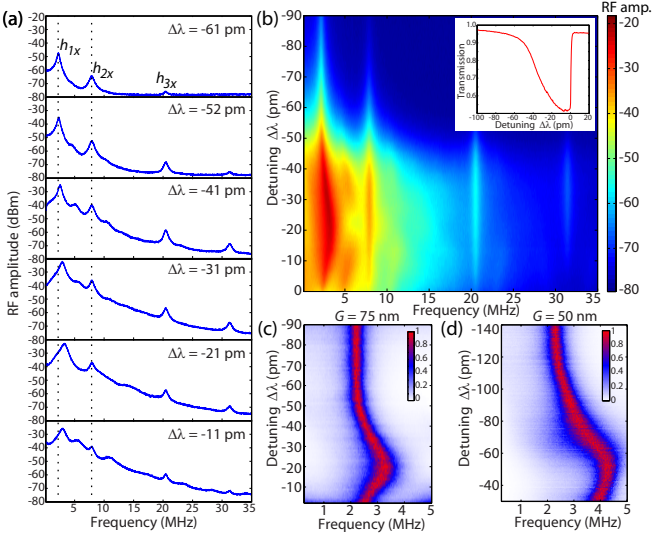}}
 \caption{(a) RF spectra from a device ($W$=65 nm, $G$=75 nm)
 with $P_{in}=446$ $\mu$W at different laser-cavity detunings $\Delta\lambda$.
 (b) Image plot of the RF spectra as a function of $\Delta\lambda$.  The cavity mode
 used for transduction is shown in the inset. (c) Zoomed-in portion of
 the image plot for the $h_{1x}$ mode, showing optically-induced
 stiffening.
 (d) Zoomed-in image plot for the $h_{1x}$ mode of a $W$=65 nm, $G$=50 nm device.
 In (c)-(d), the RF spectra are displayed on a linear scale, with each spectrum normalized to the
peak amplitude for that value of $\Delta\lambda$.}
\label{fig:Figure3}
\end{figure}

\noindent\textbf{Optically-induced stiffening}

Increasing the optical power coupled into the cavity causes several
notable changes in the RF spectrum, as seen in Fig.
\ref{fig:Figure3}(a)-(b) for a $W$=65 nm, $G=75$ nm device, where
the coupled power is changed by fixing $P_{\text{in}}=446$ $\mu$W
and varying the detuning $\Delta\lambda$ between the laser and
cavity mode. First, the spectral position of the $h_{1x}$ mode
changes from $\Omega_{M}/2\pi\approx2.24$ MHz at large
$\Delta\lambda$ to $\Omega^{\prime}_{M}/2\pi\approx3.26$ MHz at
$\Delta\lambda=-21$ pm before returning to close to its original
value at near-zero $\Delta\lambda$ (Fig. \ref{fig:Figure3}(c)). One
explanation for this is the optical spring effect, an
optically-generated rigidity of the mechanical oscillator, as seen
in other
works\cite{ref:Sheard,ref:Corbitt,ref:Kippenberg_Vahala_OE,ref:Eichenfield_zipper}.
In particular, if we take the measured values for $\Omega_{M}$,
$\Omega^{\prime}_{M}$, $\Delta\lambda$, $\Gamma$, and internal
cavity energy $U$ (determined by $P_{\text{in}}$, transmission
contrast, and $\Gamma$), the value of $g_{\text{OM}}$ that best
matches the maximum frequency shift is $g_{\text{OM}}/2\pi=1.4$
GHz/nm, corresponding to a gap $G\approx60$ nm for the TE$_{2,45}$
mode. Similarly, Fig. \ref{fig:Figure3}(d) shows a shift from
$\Omega_{M}/2\pi\approx2.26$ MHz at large $\Delta\lambda$ to
$\Omega^{\prime}_{M}/2\pi\approx4.25$ MHz at $\Delta\lambda=-41$ pm,
in this case for the $h_{1x}$ mode of a $W$=65 nm, $G=$50 nm device.
This shift is consistent with $g_{\text{OM}}/2\pi=3.0$ GHz/nm,
corresponding to a gap $G\approx32$ nm for the TE$_{2,45}$ mode.
Both of these gaps are smaller than the nominal values, but are
reasonable given the variation observed in SEM images of fabricated
devices.

Along with the change in frequency, the linewidth of the $h_{1x}$
mode changes from $\Gamma_{M}/2\pi\approx410$ kHz at
$\Delta\lambda=-61$ pm to $\Gamma_{M}/2\pi\approx860$ kHz at
$\Delta\lambda=-21$ pm, indicating damping.  In addition, the
increase in RF amplitude of the $h_{mx}$ modes is accompanied by a
broad background which, in certain detuning ranges, produces peaks
in the RF spectrum not seen at lower powers and at frequencies that
are not predicted by mechanical simulations of the cantilever. The
precise nature of these effects is not understood, though a likely
cause is the interplay between free-carrier and thermal effects that
takes place in silicon microdisks as the intracavity energy is
increased. Measurements of devices with and without cantilevers
(supplementary information) show behavior consistent with previous
observation of such effects\cite{ref:Johnson_TJ}.  It should also be
noted that thermal effects have been observed to generate damping
for blue-detuned excitation in other optomechanical
systems\cite{ref:Eichenfield_zipper}.

\noindent\textbf{Cantilever engineering and outlook}

\begin{figure}[t]
\centerline{\includegraphics[width=0.9\linewidth]{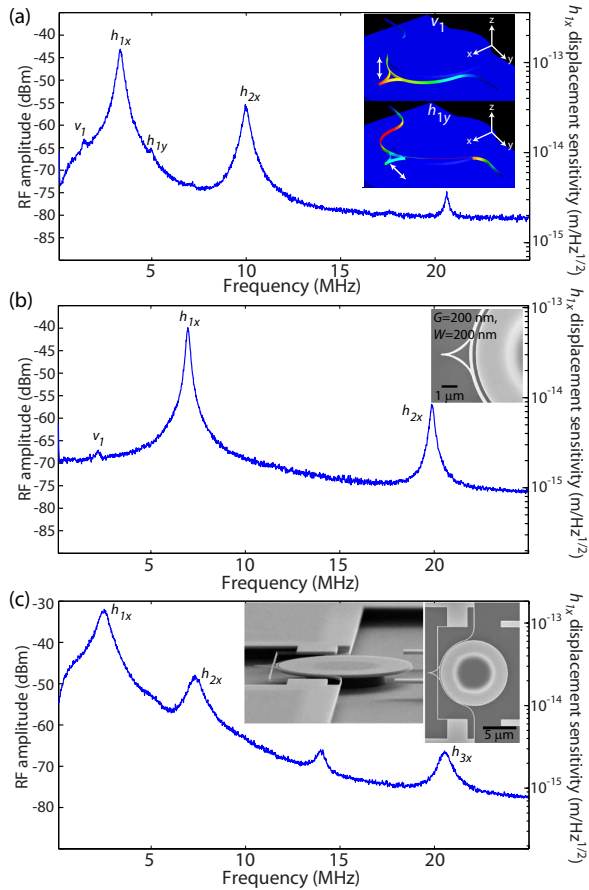}}
 \caption{(a) RF spectrum from a disk-cantilever with $W$=100 nm,
 $G$=200 nm.  Inset shows displacement profiles for cantilever
 modes not shown in Fig. \ref{fig:Figure1}(b). (b) RF spectrum from a device with $W$=200 nm,
 $G$=200 nm. (c) RF spectrum from a disk-double-cantilever with $W$=65 nm,
 $G$=50 nm. Inset shows SEMs of the device geometry.} \label{fig:Figure4}
\end{figure}

While optically-induced stiffening provides real-time control of the
cantilever properties over a certain range, a number of
modifications to its geometry can improve its applicability to
different AFM applications. The sub-N/m spring constant of the
$h_{1x}$ mode is suitable for weak force measurements in which the
cantilever is undriven, but in dynamic techniques for which the best
imaging conditions have been achieved, such as frequency modulation
AFM\cite{ref:Albrecht_FM_AFM}, spring constants in the tens of N/m
to hundreds of N/m range (or more) are desirable for small amplitude
operation\cite{ref:Giessibl_RMP}.  In our geometry, the cantilever
stiffness may be increased by increasing its width; figs.
\ref{fig:Figure4}(a)-(b) show the mechanical mode spectra for
$W$=100 nm and $W$=200 nm devices.  The $h_{1x}$ modes at
$\Omega_{M}/2\pi=3.33$ MHz and $\Omega_{M}/2\pi=6.96$ MHz agree well
with the simulated values of 3.42 MHz and 7.22 MHz. Based on the
calculated effective masses, these values correspond to a cantilever
stiffness of 0.52 N/m and 4.11 N/m, respectively, with the latter
being a 30${\times}$ increase in stiffness relative to the $h_{1x}$
mode of the $W=65$ nm device. Stiffer cantilevers can be produced by
a further increase in $W$, though degradation in the optical $Q$ is
expected unless $G$ is increased, which can then limit the
displacement sensitivity due to a reduced $g_{\text{OM}}$. Another
option is to use smaller diameter microdisks, to reduce the
cantilever length between its suspension points. Bare microdisks
have radiation-limited $Q$s $>10^6$ until their diameters are just a
couple of micrometers\cite{ref:Srinivasan12}, and simulations
predict that the $h_{1x}$ mode of a $W$=100 nm cantilever coupled to
a 4.5 $\mu$m diameter disk will occur at $7.96$ MHz $(k=1.8$ N/m).
Another important consideration is the modal structure of the
cantilever. Though we have focused on the $h_{1x}$ mode due to its
displacement profile and transduction under thermal noise, in an AFM
setting, the cantilever motion will be defined by both its actuation
mechanism and the surface it is interrogating, and its motion will
be a superposition of its modes. This includes out-of-plane ($z$
direction) and orthogonal in-plane ($y$ direction) modes such as
those seen weakly in the RF spectra of Fig.
\ref{fig:Figure4}(a)-(b). Engineering of the cantilever support
geometry can better isolate the $h_{1x}$ mode in frequency space.
The double cantilever structure shown in the SEM images of Fig.
\ref{fig:Figure4}(c) has $h_{1x}$ as its lowest frequency mode, with
the first out-of-plane mode $v_{1}$ significantly stiffened and
shifted to higher frequencies.  Figure \ref{fig:Figure4}(c) shows an
optically-transduced RF spectrum for such a device, with $W$=65 nm.
Going forward, further modifications may be made to increase the
stiffness of the cantilever, for example, through multiple short
supports to surrounding areas.  By combining this approach with
$W$=200 nm cantilevers and/or smaller diameter disks, we expect that
$k$=100 N/m devices will be achievable.  On the opposite end of the
spectrum, very soft cantilevers are also of considerable interest,
due to their application in measurements of very small
forces\cite{ref:Stowe_aN_force} such as in magnetic resonance force
microscopy\cite{ref:Rugar_single_spin}.  Reducing the cantilever
spring constant by as much as two orders of magnitude can involve
simply increasing the cantilever length and clamping it only on a
single side.

\begin{figure}[t]
\centerline{\includegraphics[width=\linewidth]{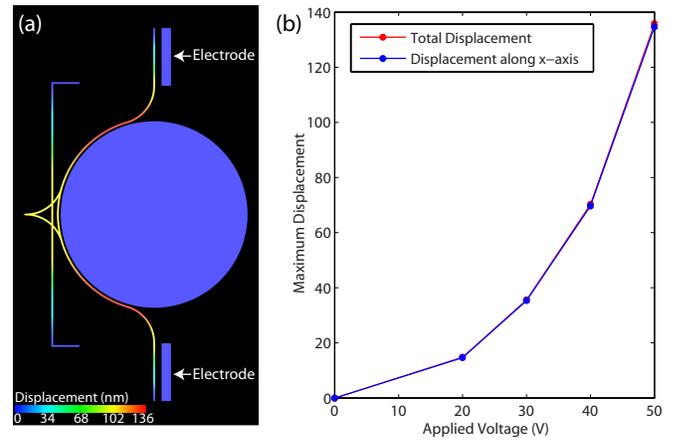}}
 \caption{(a) Simulated displacement profile for a disk-double-cantilever under
 application of 50 V to electrodes placed at the sides of the cantilevers. (b) Calculated maximum overall
 displacement and displacement along the $x$-axis as a function of applied voltage.} \label{fig:Figure5}
\end{figure}

Future dynamic AFM measurements will require an actuation mechanism
for driving the cantilever's motion.  As an illustration of one
approach, we consider electrostatic actuation through a pair of
fixed electrodes that are placed 350 nm to the side of the double
cantilever geometry (Fig. \ref{fig:Figure5}). Finite element
modeling shows that stable, steady-state displacements in excess of
100 nm can be achieved with an applied voltage under 50 V. We also
note that the displacement is primarily ($>99~\%$) along the
$x$-axis, confirming the effectiveness of the cantilever mode
engineering described above. In practice, applications such as
frequency modulation AFM will require much smaller displacements,
and regardless, the maximum detectable displacement under the
current scheme is $\approx4$ nm. This displacement level should be
achievable for an applied voltage near 5 V.  We can then estimate
the performance of this system in a frequency modulation AFM scheme
using the results from Ref. \onlinecite{ref:Albrecht_FM_AFM}, along
with the $h_{1x}$ mode frequency $\Omega_{M}/2\pi$=2.48 MHz and
linewidth $\Gamma_{M}/2\pi$=603$\pm$4 kHz for the device of Fig.
\ref{fig:Figure4}(c). The minimum detectable force $F_{\text{min}}$
and force gradient ${\delta}F_{\text{min}}^{\prime}$ are
$F_{\text{min}}=\sqrt{(4kk_{B}TB)/(\Omega_{M}Q_{M})}$=5.1$\times$10$^{-14}$
N and
${\delta}F_{\text{min}}^{\prime}=\sqrt{(4kk_{B}TB)/(\Omega_{M}Q_{M}A^2)}$=1.2$\times$10$^{-5}$
N/m, where $A$ is the cantilever oscillation amplitude (4 nm), and
$B$ is the measurement bandwidth, taken to be 50 Hz for comparison
to other experiments\cite{ref:Torbrugge}. Despite operating in an
ambient environment with $Q_{\text{M}}\approx4$, the estimated
$F_{\text{min}}$ and ${\delta}F_{\text{min}}^{\prime}$ values are
competitive with a range of systems operated in an ultra-high-vacuum
environment. In particular, silicon
cantilevers\cite{ref:Giessibl_RMP} with $k$=2 N/m,
$\Omega_{M}/2\pi$=75 kHz, and $Q_{M}=1.0{\times}10^5$ have achieved
$F_{\text{min}}=5.9{\times}10^{-15}$ N and
${\delta}F_{\text{min}}^{\prime}=3.0{\times}10^{-7}$ N/m, while
quartz tuning forks in the qPlus
configuration\cite{ref:Giessibl_QPLus} with $k$=1800 N/m,
$\Omega_{M}/2\pi$=20 kHz, and $Q_{M}=2.5{\times}10^3$ have achieved
$F_{\text{min}}=2.1{\times}10^{-12}$ N and
${\delta}F_{\text{min}}^{\prime}=1.1{\times}10^{-2}$ N/m.  More
recently\cite{ref:Torbrugge}, ultra-stiff piezoelectric quartz
length-extension resonators with $k$=$5.4{\times}10^5$ N/m,
$\Omega_{M}/2\pi$=1 MHz, and $Q_{M}=2.5{\times}10^4$ have achieved
$F_{\text{min}}=1.6{\times}10^{-12}$ N and
${\delta}F_{\text{min}}^{\prime}=8.3{\times}10^{-3}$ N/m. The
disk-cantilever system demonstrated here operates in an attractive
region of parameter space that differs from the above sensors, in
combining a MHz oscillation frequency with a 0.1 N/m to 10 N/m
stiffness (with stiffer geometries potentially feasible).

In summary, we have demonstrated sensitive transduction of the
motion of a nanoscale cantilever using a high quality factor
microdisk cavity fabricated on the same device layer. Future work
will be aimed at understanding the capabilities of this system in
AFM measurements. This will include measurements under vacuum to
determine ultimate mechanical $Q_{M}$s of the devices, and to
ascertain whether effects such as optical cooling and regenerative
oscillations\cite{ref:Kippenberg_Vahala_OE} are accessible.
Functional devices for AFM will be fabricated to expose the probe
tips to allow close proximity to other surfaces, and will be fully
integrated systems combining electrostatic or optical actuation with
on-chip resonators and waveguides\cite{ref:Wu_Solgaard_Ford}.

\noindent\textbf{Methods}

\noindent\textbf{Device Fabrication} Devices were created in a
silicon-on-insulator wafer with a 260 nm thick device layer, 1
$\mu$m thick buried oxide layer, and specified device layer
resistivity of 13.5-22.5 ohm-cm (p-type).  Fabrication steps
included electron-beam lithography of a $400$ nm-thick positive-tone
resist, an SF$_6$/C$_4$F$_8$ inductively-coupled plasma reactive ion
etch through the silicon device layer, a stabilized
H$_{2}$SO$_{4}$/H$_{2}$O$_{2}$ etch to remove the remnant resist and
other organic materials, an HF wet etch to undercut the devices and
release the cantilevers, and a critical point dry to
 finish the processing. The etch time required to go through the silicon device layer is a
function of cantilever-disk gap, with an $\approx30\%$ increase in
etch time required for $G=50$ nm devices relative to $G=200$ nm
devices.

\noindent\textbf{Device Simulation} Mechanical eigenfrequencies and
eigenmodes of the cantilever and disk were studied using a
commercial finite element software package. Silicon was modeled as
an elastic cubic material using three independent elastic
constants\cite{ref:Senturia} with (100) orientation, and clamped
boundary conditions were assumed at the cantilever ends. Mesh
refinement studies indicate that numerical errors are below the
uncertainty resulting from imperfect knowledge of the cantilever
geometry, which is generally a few percent of the reported values.
For the reported mode frequencies $\Omega_{M}$ and effective masses
$m$, zero residual stress was assumed. In a separate numerical
study, all cantilever mode frequencies were shown to be
approximately independent (within a few percent) of the residual
stress for stress values under $\pm$100 MPa. The mode stiffness was
calculated as $k=m\Omega_{M}^2$.

Electrostatic actuation was modeled by iteratively solving a coupled
three-dimensional static-mechanical problem and a three-dimensional
electrostatic problem. The former fixes the elastic properties and
clamped boundary conditions at the four double-cantilever ends to be
the same. The mechanically fixed electrodes are 260 nm thick, 500 nm
wide, and 3 $\mu$m long, and the gap between them and the cantilever
is 350 nm. The same fixed voltage is applied to both electrodes
(doped silicon is assumed to be a perfect conductor), while the
cantilever is assumed to be at the ground potential. Given the
applied voltages and shape of the deformable cantilever and fixed
electrodes, the electrostatic force densities on all cantilever
surfaces are calculated using a boundary element method. The
calculated force densities were then applied as boundary conditions
and the mechanical problem was solved to find the new deformed beam
shape. The electrostatic and mechanical solvers were iterated until
the solution converged to a stable value for each applied voltage.
Mesh refinement studies were conducted on the electrostatic surface
mesh to ensure numerical accuracy. The microdisk and substrate were
assumed to be at ground potential and not included in this model for
simplicity. This is justified because the cantilever-electrode gap
is much smaller than the distances between the electrodes and either
the microdisk or substrate.

Optical eigenfrequencies and eigenmodes of the disk-cantilever
system were found numerically using a second commercial finite
element software package. The silicon layer was modeled as having an
index of refraction $n$=3.4 surrounded by air ($n$ =1), and both
materials were assumed lossless and non-magnetic. The model size was
chosen to be large enough to fully contain the modes studied, with
scattering boundary conditions on the outside surfaces.  A mesh
refinement study was conducted to ensure numerical accuracy.
$g_{\text{OM}}$ for the $h_{1x}$ mechanical mode as a function of
the gap $G$ was obtained by linearly translating the cantilever with
respect to the disk along the $x$-axis from the initial
cantilever-disk gap $G$=100 nm. For each value of $G$ between 30 nm
and 300 nm the cantilever was further deformed using the calculated
$h_{1x}$ mode shape. The modal deformations were 0 and ${\pm}d$,
where $d$ varied from 2 nm for $G=30$ nm to 10 nm for $G>$100 nm.
For each $G$ and deformation the frequencies and $Q$s for multiple
optical modes were numerically calculated. For each optical mode and
$G$ the derivative of the frequency with respect to modal
deformation was obtained using the slope of a linear fit. In all
cases, the gap changes and cantilever deformations were implemented
by numerically deforming the same original mesh to obtain the
desired cantilever shape and position before solving the optical
eigenvalue problem.

\noindent\textbf{Device Characterization} Devices were characterized
using a swept-wavelength external cavity tunable diode laser with a
time-averaged linewidth $<90$ MHz and absolute stepped wavelength
accuracy of $\pm1$ pm. The wavelength tuning range and linearity are
calibrated using an acetylene reference cell, so that the
uncertainty in optical cavity $Q$s is dominated by fits to the data.
Light is coupled into and out of the cavities using an optical fiber
taper waveguide in a $N_{2}$-purged environment at atmospheric
pressure and room temperature.  Cavity transmission spectra were
recorded using a variable gain InGaAs photoreceiver with a typical
bandwidth of 775 kHz, noise equivalent power (NEP) of 1.25
pW/$\sqrt{\text{Hz}}$, and gain of 4.5${\times}10^4$ V/W.  RF
spectra were recorded using either a 0 MHz (DC) to 125 MHz InGaAs
photoreceiver (NEP=2.5 pW/$\sqrt{\text{Hz}}$, gain=4${\times}10^4$
V/W) or DC to 1.1 GHz InGaAs avalanche photodiode (NEP=1.6
pW/$\sqrt{\text{Hz}}$, gain=1.4${\times}10^4$ V/W) whose output was
sent into a 9kHz to 3.0 GHz electronic spectrum analyzer with
resolution bandwidth typically set at 30 kHz.  RF frequencies and
linewidths are determined by Lorentzian fits to the data, with
uncertainties given by the 95 $\%$ confidence intervals of the fit
(uncertainties are not written if they are smaller than the number
of digits to which the value is quoted). Optical frequencies and
linewidths are determined by a least squares fit to the data using a
doublet model that takes into account both clockwise and
counterclockwise whispering gallery modes and their coupling due to
backscattering\cite{ref:Borselli2}.

\noindent\textbf{Phase modulator calibration} As a consistency check
on the calibration of displacement
sensitivity\cite{ref:Schliesser_NJP}, we use an electro-optic phase
modulator (Fig. \ref{fig:Figure2}(a)) of known modulation depth
$\delta\phi$ and frequency $\Omega_{mod}$ to generate a tone in the
RF spectrum, at 44 MHz in Fig. \ref{fig:Figure2}(e).  This
modulation peak is equivalent to an effective mechanical oscillation
amplitude $x_{mod}=\delta\phi(\omega_{mod}/g_{\text{OM}})$, and can
provide a check on $x_{rms}$, but is limited by the accuracy to
which $g_{\text{OM}}$ is known. For Fig. \ref{fig:Figure2}, assuming
$G=100$ nm and that the optical mode used for transduction is the
TE$_{2,45}$ mode, $g_{\text{OM}}/2\pi=0.61$ GHz/nm produces a value
$x\approx192$ pm that is $\approx20$ $\%$ greater than $x_{rms}=160$
pm. A likely source for the discrepancy is imperfect knowledge of
the gap; for example, a 10 nm decrease in it would completely
account for the difference between the two values.

\noindent \textbf{Acknowledgements} We thank Lei Chen of the CNST
NanoFab for assistance with silicon etch development.

\noindent \textbf{Author Contributions} K.S. led device
characterization, H.M. led device fabrication, and V.A. led device
simulation, while M.T.R. and M.D. helped build the optical
characterization setup.  All authors contributed to preparation of
the manuscript.

\noindent \textbf{Author Information} The authors declare no
competing financial interests.  Correspondence and requests for
material should be addressed to K.S. and V.A.

\clearpage

\newpage

\appendix{\bf{Supplementary Information}}
\setcounter{figure}{0}

\subsection{Optical cavity modes and optomechanical coupling}
Finite-element method (FEM) simulations indicate that the $p$=1 and
$p$=2 modes of TE polarization and $p$=1 modes of TM polarization
have high $Q$s ($>10^5$) for sufficiently thin cantilevers.
Simulation results for $W$=65 nm cantilevers are shown in Fig.
\ref{fig:SFigure1}(a).  Similar simulations for $W$=100 nm
cantilevers indicate a reduction in $Q$ by as much as a factor of 3,
though it nevertheless remains above $10^5$.  As discussed in the
text, while most fabricated devices have cavity $Q$s in the range of
$5{\times}10^4$ to $2{\times}10^5$, a few exhibit $Q$s as high as
$\approx6{\times}10^5$ (Fig. \ref{fig:SFigure1}(b)). The optically
transduced RF spectra in such devices often show a strong amplitude
for not only the $h_{1x}$ modes, but also $h_{my}$ and $v_{n}$
modes.  This suggests some amount of asymmetry in the cantilever
structure not found in the majority of the devices (such as those
studied in the main text).

\makeatletter \renewcommand{\thefigure}{S\@arabic\c@figure}
\begin{figure}[t]
\centerline{\includegraphics[width=\linewidth]{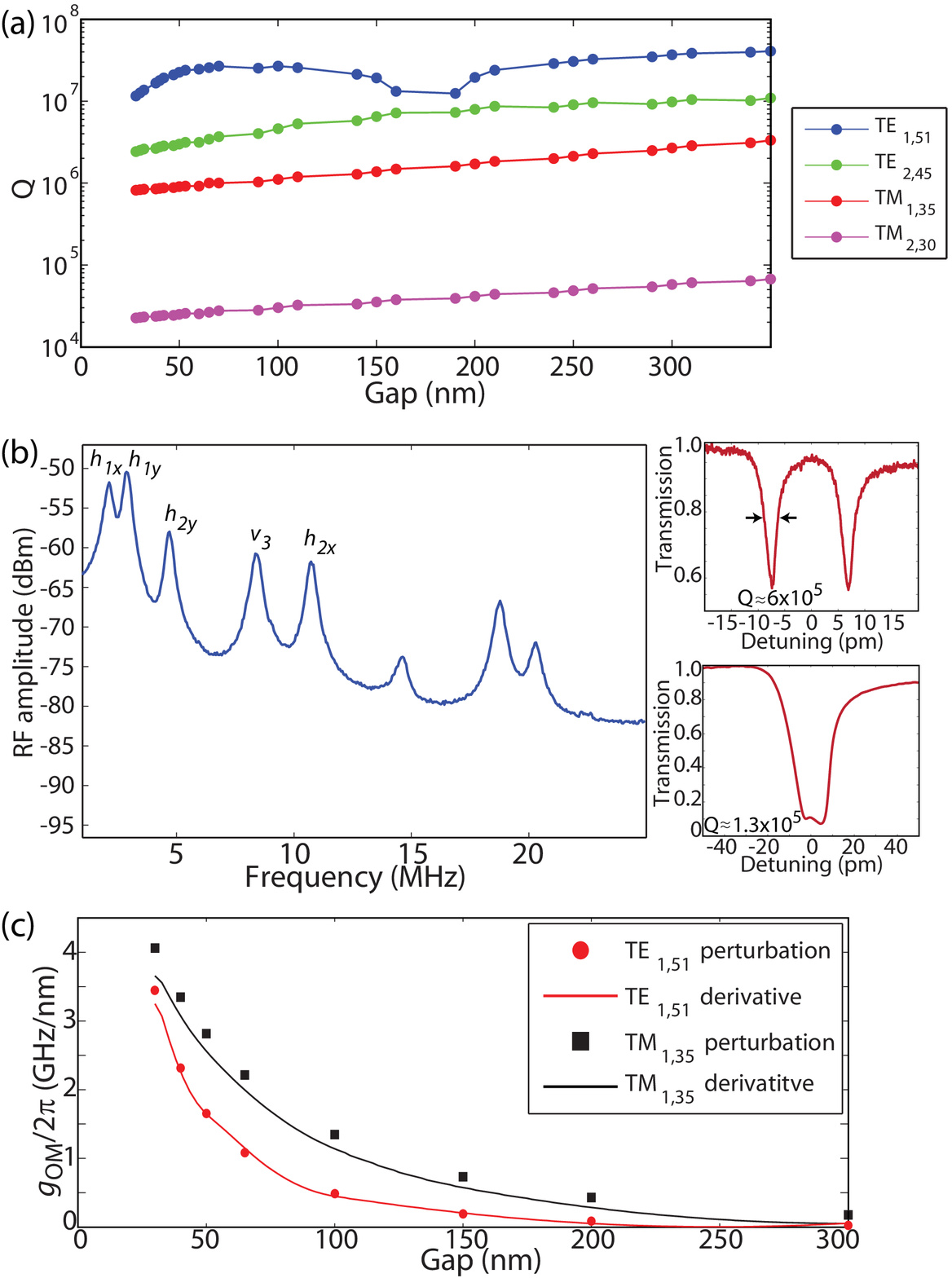}}
 \caption{(a) FEM-calculated optical $Q$s for TE and TM polarized modes of a disk-cantilever with $W$=65 nm.
 (b) Thermal noise spectrum of a $G$=100 nm, $W$=100 nm disk-cantilever device.  To the right are
 two optical modes of the structure; the bottom scan is of the mode used in
 transduction. (c) Predicted optomechanical coupling $g_{\text{OM}}$ between the $h_{1x}$
cantilever mode and 1st order radial TE and TM modes of the
microdisk. Points are from a series of simulations for varying $G$,
while solid lines are calculated using the perturbation theory
method described in Ref. S1} \label{fig:SFigure1}
\end{figure}

Generally, the measured optical $Q$s decrease with decreasing gap
and increasing cantilever width.  Smaller gaps can also be
problematic because the time required to etch through the silicon
layer goes up as the gap size is reduced, potentially leading to
mask erosion and a roughening of the disk sidewalls. The
optomechanical coupling $g_{\text{OM}}$, on the other hand,
increases with decreasing gap and increasing cantilever width. The
calculated $g_{\text{OM}}$ for $p$=1 modes with a $W$=100 nm
cantilever is shown in Fig. \ref{fig:SFigure1}(c), and can be
$\approx25$ $\%$ larger than the values calculated for $W$=65 nm in
Fig. \ref{fig:Figure1}(d).

\subsection{Hansch-Couillaud polarization spectroscopy}

For future experiments (including AFM applications) it will likely
be necessary to lock the probe laser to the cavity. This can be done
by beating the signal exiting the cavity with a strong local
oscillator (LO), thereby measuring phase fluctuations due to
cantilever motion and giving access to a dispersive signal needed
for locking.  A particularly convenient approach, Hansch-Couillaud
polarization spectroscopy as described in Refs. S2 and S3 and shown
schematically in Fig. \ref{fig:SFigure2}(a), sets the polarization
so that only part of the input field couples to the cavity, with the
orthogonal polarization serving as the LO. The interference signal
is analyzed using a $\lambda/4$ waveplate and polarizing beam
splitter, whose outputs are measured on a 100 MHz balanced
photodetector.  The difference signal produced by scanning the laser
over a cavity resonance is shown in the inset to Fig.
\ref{fig:SFigure2}(b). Positioning the laser on resonance and
measuring the RF fluctuations in this signal produces the thermal
noise spectrum shown in Fig. \ref{fig:SFigure2}(b).

\makeatletter \renewcommand{\thefigure}{S\@arabic\c@figure}
\begin{figure}[t]
\centerline{\includegraphics[width=\linewidth]{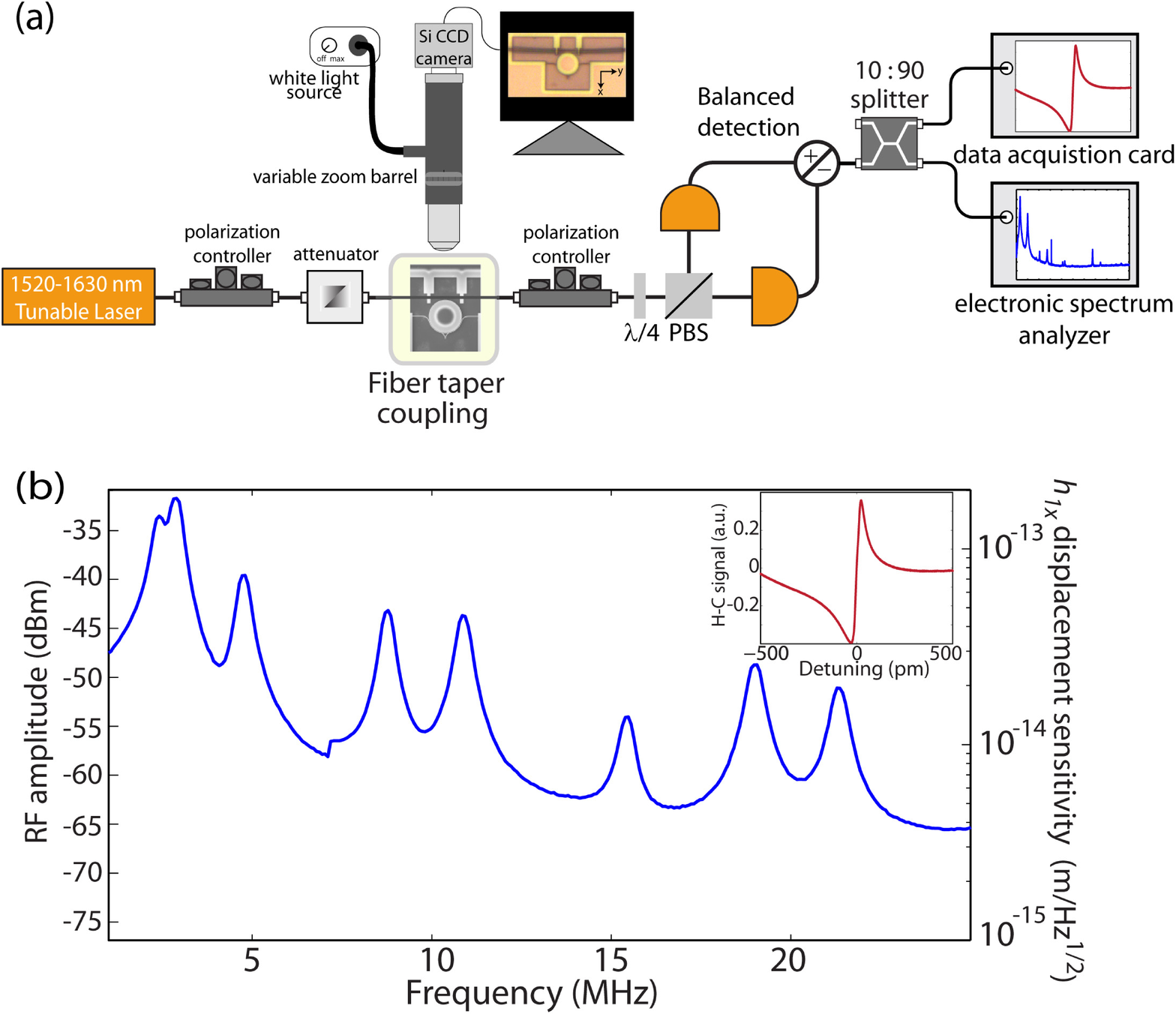}}
 \caption{(a) Schematic of the setup used for Hansch-Couillaud
 homodyne spectroscopy.  The input polarization into the cavity is
 set so that a small fraction of the signal is coupled into the mode
 of interest, while the remainder acts as a local oscillator.  (b) RF spectrum measured using Hansch-Couillaud homodyne
spectroscopy.  Inset shows the difference signal produced when the
laser is scanned over the cavity resonance.} \label{fig:SFigure2}
\end{figure}

\subsection{Self-induced optical modulation and free carrier effects}

Two-photon absorption is well-known to play an important role in
silicon nanophotonics$^\text{S4}$, with the subsequent generation of
phonons and free carriers giving rise to both optical dispersion and
loss, and with the associated lifetimes affecting the speed of
devices intended to exploit these effects.  In Ref. S5, Johnson
\textit{et al.} observed that under sufficiently strong continuous
wave input, silicon microdisks of similar dimensions to those
studied in this work exhibited steady-state oscillations in their
transmitted power.  The authors attributed this to competing thermal
and free-carrier effects, as the dispersion in the refractive index
caused by the two effects are opposite in sign (red-shift for
thermal, blue-shift for free carriers), and as the cavity mode
position shifts due to this change in refractive index, the
circulating power in the cavity changes, thereby changing the rate
at which heat and free carriers are created.  Looking in the
frequency domain, the RF spectrum of the transmitted signal
displayed a number of sharp peaks with a spacing of a few hundred
kHz.  We have observed similar phenomena in our bare (no cantilever)
microdisks.  Fig. \ref{fig:SFigure3}(a) shows both a broad (up to
200 MHz) and zoomed-in (up to 15 MHz) spectrum of the transmitted
signal from a microdisk with a $Q\approx3{\times}10^5$ mode coupled
to by a fiber taper waveguide with $P_{\text{in}}\approx$440 $\mu$W
at 1533.6 nm.  A comb of sharp peaks is observed in the RF spectrum,
with a nearest-neighbor spacing that is typically $\approx3.23$ MHz.

\makeatletter \renewcommand{\thefigure}{S\@arabic\c@figure}
\begin{figure}[t]
\centerline{\includegraphics[width=\linewidth]{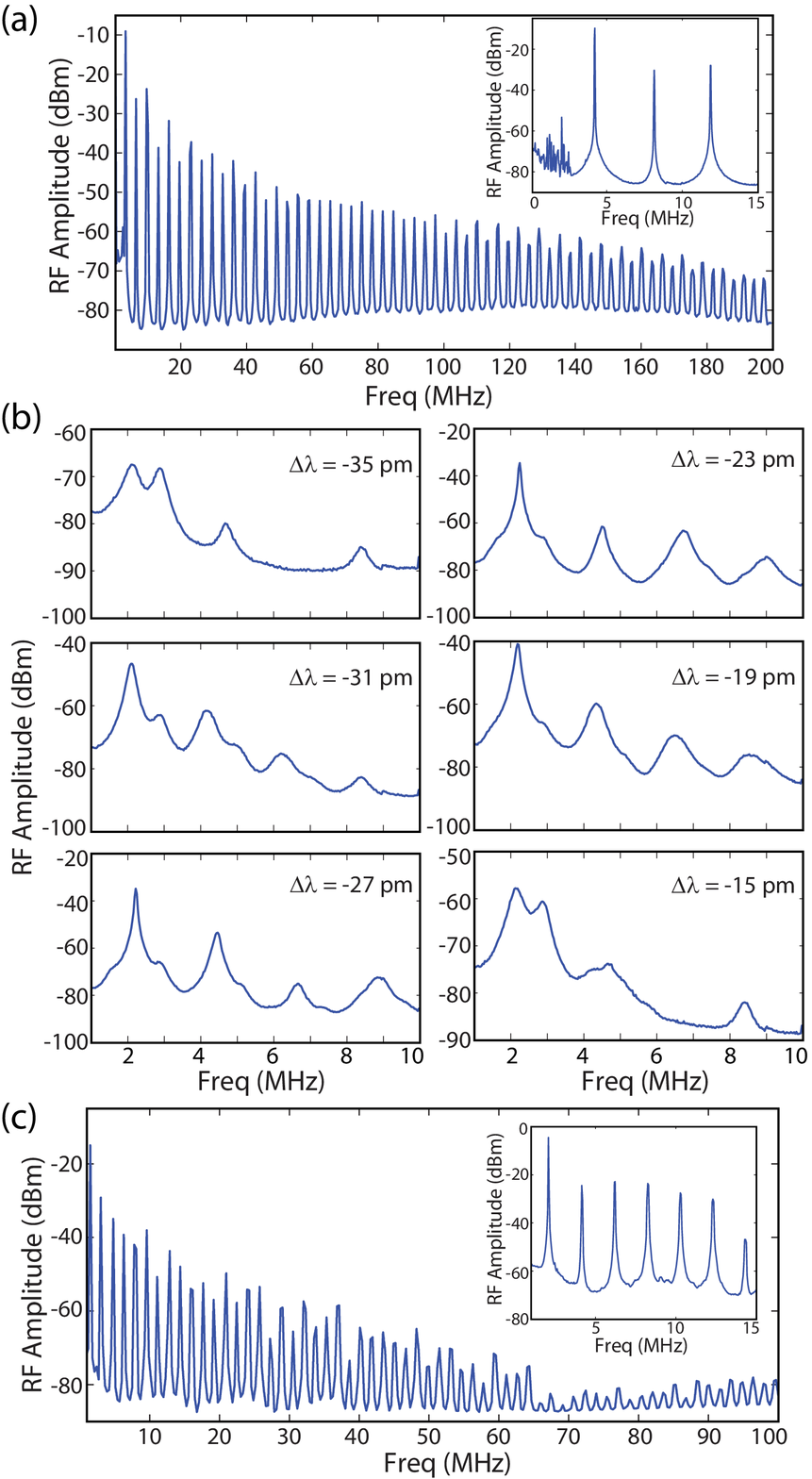}}
 \caption{(a) Broad and zoomed-in RF spectrum of a bare microdisk (no cantilever)
 with $P_{\text{in}}=440$ $\mu$W coupled to a $Q\approx3{\times}10^5$
cavity mode. (b) RF spectra from the cantilever-microdisk system of
Fig. \ref{fig:SFigure1} as a function of laser-cavity detuning
$\Delta\lambda$ with $P_{\text{in}}=60$ $\mu$W coupled to a
$Q\approx1.3{\times}10^5$ cavity mode. (c) RF spectrum of the
cantilever-microdisk system with $P_{\text{in}}\approx1400$ $\mu$W.
Inset is a zoomed in high-resolution scan of a portion of the
spectrum.} \label{fig:SFigure3}
\end{figure}

The devices shown in the main body of the text have somewhat lower
optical $Q$s than the above device, which likely explains why
similarly sharp RF peaks are not observed at similar input powers.
Instead, the RF spectra (Fig. \ref{fig:Figure3}(a)) look to be a
superposition of the spectrum due to mechanical oscillations of the
cantilever and/or disk and the spectrum due to competing free
carrier and thermal effects within the disk, albeit below the
threshold for oscillation.  To consider this point further, we look
at the disk-cantilever device of Fig. \ref{fig:SFigure1}(b),
coupling to a TE$_{2,n}$ mode with $Q\approx1.3{\times}10^5$ (bottom
inset of Fig. \ref{fig:SFigure2}(b)). If we initially restrict
$P_{\text{in}}\approx60$ $\mu$W and vary the laser-cavity detuning
$\Delta\lambda$, we generate fig. \ref{fig:SFigure3}(b). We see that
for initial large detunings ($\Delta\lambda=-35$ pm), the RF
spectrum is dominated by the mechanical modes of the cantilever, but
as the detuning decreases, a background with broad resonances is
superimposed ($\Delta\lambda=-31$ pm) and dominates
($\Delta\lambda=-23$ pm), before eventually the mechanical modes
re-appear for small enough detunings ($\Delta\lambda=-15$ pm). It is
believed that the broad background is due to the same
thermal/free-carrier effects seen in Ref. S5 and in the bare disk of
Fig. \ref{fig:SFigure3}(a).  Indeed, if $P_{\text{in}}$ is increased
to a few hundred $\mu$W, a qualitatively similar RF spectrum (Fig.
\ref{fig:SFigure3}(c)) is observed - here the mechanical modes are
completely dominated by the thermal/free-carrier effects.

Quantitative modeling of this behavior can be accomplished in a
manner similar to that of Ref. S6, where the equations of bare
optomechanics (evolution of the intracavity optical field amplitude
and mechanical position) were augmented by an equation for the
cavity temperature increase.  We now have to add a fourth
differential equation, to account for the change in free carrier
population.  Following the treatment of thermal and free-carrier
terms presented in Ref. S5, we have:

%\begin{subequations}
%\begin{align*}
%\label{a_eqn}
%&\frac{da}{dt}=-\frac{1}{2}\Biggl(\Gamma+\frac{\Gamma_{\text{TPA}}\beta_{\text{Si}}c^2}{V_{\text{TPA}}n_{g}^2}|a(t)|^2+\frac{\sigma_{\text{Si}}c}{n_{g}}N(t)\Biggr)a(t)
%\\ & \quad
%+i\Biggl(\delta\omega_{o}+g_{\text{OM}}x+\frac{\omega_{o}}{n_{\text{Si}}}\frac{dn_{\text{Si}}}{dT}{\Delta}T(t)+\frac{\omega_{o}}{n_{\text{Si}}}\frac{dn_{\text{Si}}}{dN}N(t)\Biggr)a(t) + {\kappa}s\\
%\label{x_eqn}
%&\frac{dx}{dt}=-\Gamma_{M}\frac{dx}{dt}-\Omega^2_{M}x-\frac{|a(t)|^2g_{\text{OM}}}{\omega_{c}m}
%\\
%\label{N_eqn}
%&\frac{dN}{dt}=-\gamma^{\prime}_{\text{fc}}N(t)+\frac{\Gamma_{\text{FCA}}\beta_{\text{Si}}c^2}{2\hbar\omega_{p}n_{g}^2}\frac{|a(t)|^4}{V_{\text{FCA}}^2}
%\\
%\label{thermal}
%&\frac{d\Delta
%T}{dt}=-\gamma_{\text{th}}{\Delta}T(t)+\frac{\Gamma_{\text{disk}}}{\rho_{\text{Si}}c_{\text{p,Si}}V_{\text{disk}}}\Biggl(\gamma_{\text{lin}}+\frac{\Gamma_{\text{TPA}}\beta_{\text{Si}}c^2}{V_{\text{TPA}}n_{g}^2}|a(t)|^2
%\\ & \quad
%+\frac{\sigma_{\text{Si}}c}{n_{g}}N(t)\Biggr)|a(t)|^2.
%\end{align*}
%\end{subequations}

\begin{subequations}
\begin{align*}
%\label{a_eqn}
&\frac{da}{dt}=-\frac{1}{2}\Biggl(\Gamma+\alpha_{\text{TPA}}|a(t)|^2+\beta_{\text{FCA}}N(t)\Biggr)a(t)
\\ & \quad
+i\Biggl(\delta\omega_{c}+g_{\text{OM}}x+g_{th}{\Delta}T(t)+g_{fc}N(t)\Biggr)a(t) + {\kappa}s\\
%\label{x_eqn}
&\frac{dx}{dt}=-\Gamma_{M}\frac{dx}{dt}-\Omega^2_{M}x-\frac{|a(t)|^2g_{\text{OM}}}{\omega_{c}m}
\\
%\label{thermal}
&\frac{d\Delta
T}{dt}=-\gamma_{\text{th}}{\Delta}T(t)+c_{th}\Biggl(\Gamma_{abs}+\alpha_{\text{TPA}}|a(t)|^2
+\beta_{\text{FCA}}N(t)\Biggr)|a(t)|^2
%\label{N_eqn}
\\
&\frac{dN}{dt}=-\gamma^{\prime}_{\text{fc}}N(t)+\chi_{\text{FCA}}|a(t)|^4
\end{align*}
\end{subequations}

\noindent where for simplicity we have assumed a single-mode cavity
- a more detailed treatment would include both modes of the
microdisk and their coupling via backscattering.  The first equation
describes the intracavity field amplitude $a(t)$, where the first
term on the right is its decay due to intrinsic and waveguide loss
$\Gamma$, two-photon absorption ($\alpha_{\text{TPA}}$) and
free-carrier absorption ($\beta_{\text{FCA}}$), while the second
term includes the laser-cavity detuning $\delta\omega_{c}$ and
dispersion due to optomechanical coupling ($g_{\text{OM}}$),
thermo-optic tuning ($g_{th}$), and free-carrier dispersion
($g_{fc}$).  The second equation describes the mechanical motion
$x(t)$ with frequency $\Omega_{M}$ and damping $\Gamma_{M}$ and
driven by the coupling to the optical field. The third equation
describes the cavity temperature change ${\Delta}T(t)$, where the
cavity has a heat capacity $c_{th}$, the temperature decays with a
rate $\gamma_{th}$, and is generated by linear absorption
($\Gamma_{abs}$ is the portion of total optical loss that
contributes), two-photon absorption ($\alpha_{\text{TPA}}$), and
free-carrier absorption ($\beta_{\text{FCA}}$). Finally, the fourth
equation describes the modal free carrier population $N(t)$, which
decays at a rate $\gamma^{\prime}_{\text{fc}}$ and is generated in
proportion to the square of intracavity energy, with proportionality
$\chi_{\text{FCA}}$.  The various coefficients in the above
equations are described in detail in Ref. S5, and are a combination
of physical properties such as the two-photon absorption coefficient
of silicon and the absorption cross-section for free carriers, as
well as cavity mode properties such as its group index and different
confinement factors and modal volumes weighted according to the
electric-field dependence of the given process (e.g., two-photon
absorption or free-carrier absorption).

An analysis of the above equations will produce correction terms to
the mechanical frequency $\Omega_{M}$ and linewidth $\Gamma_{M}$,
and may help provide a better understanding of the power-dependent
RF spectra shown in the main text (Fig. \ref{fig:Figure3}).  For
example, Eichenfield \textit{et.al}$^\text{S6}$ determined that in
their SiN$_{x}$ photonic crystal nanobeam devices, heating
significantly affected the linewidth as a function of detuning, but
not the mechanical frequency.  In comparison, in addition to
linewidth modification (observation of damping for blue-detuned
excitation), the frequency dependence on detuning for our devices
(Fig. \ref{fig:Figure3}(c)-(d)) does appear to show some effect, in
that the shape of the curves near zero-detuning is not nearly as
sharply-sloped as the equations of bare optomechanics predict.

\subsection{References}

\noindent S1. \bibinfo{author}{{Eichenfield}, M.},
\bibinfo{author}{{Chan}, J.},
  \bibinfo{author}{{Camacho}, R.~M.}, \bibinfo{author}{{Vahala}, K.~J.} \&
  \bibinfo{author}{{Painter}, O.}
\newblock \bibinfo{title}{{Optomechanical crystals}}.
\newblock \emph{\bibinfo{journal}{Nature}} \textbf{\bibinfo{volume}{462}},
  \bibinfo{pages}{78--82} (\bibinfo{year}{2009})

\noindent S2. \bibinfo{author}{{Schliesser}, A.},
\bibinfo{author}{{Rivi{\`e}re}, R.},
  \bibinfo{author}{{Anetsberger}, G.}, \bibinfo{author}{{Arcizet}, O.} \&
  \bibinfo{author}{{Kippenberg}, T.~J.}
\newblock \bibinfo{title}{{Resolved-sideband cooling of a micromechanical
  oscillator}}.
\newblock \emph{\bibinfo{journal}{Nature Physics}}
  \textbf{\bibinfo{volume}{4}}, \bibinfo{pages}{415--419}
  (\bibinfo{year}{2008}).

\noindent S3. \bibinfo{author}{{Schliesser}, A.},
\bibinfo{author}{{Anetsberger}, G.},
  \bibinfo{author}{{Rivi{\`e}re}, R.}, \bibinfo{author}{{Arcizet}, O.} \&
  \bibinfo{author}{{Kippenberg}, T.~J.}
\newblock \bibinfo{title}{{High-sensitivity monitoring of micromechanical
  vibration using optical whispering gallery mode resonators}}.
\newblock \emph{\bibinfo{journal}{New Journal of Physics}}
  \textbf{\bibinfo{volume}{10}}, \bibinfo{pages}{095015}
  (\bibinfo{year}{2008}).

\noindent S4. \bibinfo{author}{{Lin}, Q.},
\bibinfo{author}{{Painter}, O.~J.} \&
  \bibinfo{author}{{Agrawal}, G.~P.}
\newblock \bibinfo{title}{{Nonlinear optical phenomena in silicon waveguides:
  modeling and applications}}.
\newblock \emph{\bibinfo{journal}{Opt. Express}} \textbf{\bibinfo{volume}{15}},
  \bibinfo{pages}{16604--16644} (\bibinfo{year}{2007}).

\noindent S5. \bibinfo{author}{{Johnson}, T.~J.},
\bibinfo{author}{{Borselli}, M.} \&
  \bibinfo{author}{{Painter}, O.}
\newblock \bibinfo{title}{{Self-induced optical modulation of the transmission
  through a high-Q silicon microdisk resonator}}.
\newblock \emph{\bibinfo{journal}{Opt. Express}} \textbf{\bibinfo{volume}{14}},
  \bibinfo{pages}{817--831} (\bibinfo{year}{2006}).

\noindent S6. \bibinfo{author}{{Eichenfield}, M.},
\bibinfo{author}{{Camacho}, R.},
  \bibinfo{author}{{Chan}, J.}, \bibinfo{author}{{Vahala}, K.~J.} \&
  \bibinfo{author}{{Painter}, O.}
\newblock \bibinfo{title}{{A picogram- and nanometre-scale photonic-crystal
  optomechanical cavity}}.
\newblock \emph{\bibinfo{journal}{Nature}} \textbf{\bibinfo{volume}{459}},
  \bibinfo{pages}{550--555} (\bibinfo{year}{2009}).

% ****** End of file apssamp.tex ******


\begin{thebibliography}{10}
\expandafter\ifx\csname url\endcsname\relax
  \def\url#1{\texttt{#1}}\fi
\expandafter\ifx\csname urlprefix\endcsname\relax\def\urlprefix{URL
}\fi \providecommand{\bibinfo}[2]{#2}
\providecommand{\eprint}[2][]{\url{#2}}

\bibitem{ref:Ekinci_Roukes_RSI}
\bibinfo{author}{{Ekinci}, K.~L.} \& \bibinfo{author}{{Roukes}, M.~L.}
\newblock \bibinfo{title}{{Nanoelectromechanical systems}}.
\newblock \emph{\bibinfo{journal}{Review of Scientific Instruments}}
  \textbf{\bibinfo{volume}{76}}, \bibinfo{pages}{061101}
  (\bibinfo{year}{2005}).

\bibitem{ref:Li_Tang_Roukes}
\bibinfo{author}{{Li}, M.}, \bibinfo{author}{{Tang}, H.~X.} \&
  \bibinfo{author}{{Roukes}, M.~L.}
\newblock \bibinfo{title}{{Ultra-sensitive NEMS-based cantilevers for sensing,
  scanned probe and very high-frequency applications}}.
\newblock \emph{\bibinfo{journal}{Nature Nanotechnology}}
  \textbf{\bibinfo{volume}{2}}, \bibinfo{pages}{114--120}
  (\bibinfo{year}{2007}).

\bibitem{ref:Craighead_mass_sensing}
\bibinfo{author}{{Ilic}, B.} \emph{et~al.}
\newblock \bibinfo{title}{{Attogram detection using nanoelectromechanical
  oscillators}}.
\newblock \emph{\bibinfo{journal}{J. Appl. Phys.}}
  \textbf{\bibinfo{volume}{95}}, \bibinfo{pages}{3694--3703}
  (\bibinfo{year}{2004}).

\bibitem{ref:Rugar}
\bibinfo{author}{{Rugar}, D.}, \bibinfo{author}{{Yannoni}, C.~S.} \&
  \bibinfo{author}{{Sidles}, J.~A.}
\newblock \bibinfo{title}{{Mechanical detection of magnetic resonance}}.
\newblock \emph{\bibinfo{journal}{Nature}} \textbf{\bibinfo{volume}{360}},
  \bibinfo{pages}{563--566} (\bibinfo{year}{1992}).

\bibitem{ref:Giessibl_RMP}
\bibinfo{author}{{Giessibl}, F.}
\newblock \bibinfo{title}{{Advances in atomic force microscopy}}.
\newblock \emph{\bibinfo{journal}{Reviews of Modern Physics}}
  \textbf{\bibinfo{volume}{75}}, \bibinfo{pages}{949--983}
  (\bibinfo{year}{2003}).

\bibitem{ref:Walters_RSI}
\bibinfo{author}{{Walters}, D.~A.} \emph{et~al.}
\newblock \bibinfo{title}{{Short cantilevers for atomic force microscopy}}.
\newblock \emph{\bibinfo{journal}{Review of Scientific Instruments}}
  \textbf{\bibinfo{volume}{67}}, \bibinfo{pages}{3583--3590}
  (\bibinfo{year}{1996}).

\bibitem{ref:Kawakatsu}
\bibinfo{author}{{Kawakatsu}, H.} \emph{et~al.}
\newblock \bibinfo{title}{{Towards atomic force microscopy up to 100 MHz}}.
\newblock \emph{\bibinfo{journal}{Review of Scientific Instruments}}
  \textbf{\bibinfo{volume}{73}}, \bibinfo{pages}{2317--2320}
  (\bibinfo{year}{2002}).

\bibitem{ref:Sahin}
\bibinfo{author}{Sahin, O.}, \bibinfo{author}{Magonov, S.},
  \bibinfo{author}{Su, C.}, \bibinfo{author}{Quate, C.~F.} \&
  \bibinfo{author}{Solgaard, O.}
\newblock \bibinfo{title}{{An atomic force microscope tip designed to measure
  time-varying nanomechanical forces}}.
\newblock \emph{\bibinfo{journal}{Nature Nanotechnology}}
  \textbf{\bibinfo{volume}{{2}}}, \bibinfo{pages}{{507--514}}
  (\bibinfo{year}{{2007}}).

\bibitem{ref:Meyer_AFM}
\bibinfo{author}{{Meyer}, G.} \& \bibinfo{author}{{Amer}, N.~M.}
\newblock \bibinfo{title}{{Novel optical approach to atomic force microscopy}}.
\newblock \emph{\bibinfo{journal}{Appl. Phys. Lett.}}
  \textbf{\bibinfo{volume}{53}}, \bibinfo{pages}{1045--1047}
  (\bibinfo{year}{1988}).

\bibitem{ref:Rugar_AFM}
\bibinfo{author}{{Rugar}, D.}, \bibinfo{author}{{Mamin}, H.~J.} \&
  \bibinfo{author}{{Guethner}, P.}
\newblock \bibinfo{title}{{Improved fiber-optic interferometer for atomic force
  microscopy}}.
\newblock \emph{\bibinfo{journal}{Appl. Phys. Lett.}}
  \textbf{\bibinfo{volume}{55}}, \bibinfo{pages}{2588--2590}
  (\bibinfo{year}{1989}).

\bibitem{ref:Arcizet}
\bibinfo{author}{{Arcizet}, O.} \emph{et~al.}
\newblock \bibinfo{title}{{High-Sensitivity Optical Monitoring of a
  Micromechanical Resonator with a Quantum-Limited Optomechanical Sensor}}.
\newblock \emph{\bibinfo{journal}{Phys. Rev. Lett.}}
  \textbf{\bibinfo{volume}{97}}, \bibinfo{pages}{133601}
  (\bibinfo{year}{2006}).

\bibitem{ref:Hoogenboom}
\bibinfo{author}{{Hoogenboom}, B.~W.} \emph{et~al.}
\newblock \bibinfo{title}{{A Fabry-Perot interferometer for micrometer-sized
  cantilevers}}.
\newblock \emph{\bibinfo{journal}{Appl. Phys. Lett.}}
  \textbf{\bibinfo{volume}{86}}, \bibinfo{pages}{074101}
  (\bibinfo{year}{2005}).

\bibitem{ref:Ekinci_interferometer}
\bibinfo{author}{{Kouh}, T.}, \bibinfo{author}{{Karabacak}, D.},
  \bibinfo{author}{{Kim}, D.~H.} \& \bibinfo{author}{{Ekinci}, K.~L.}
\newblock \bibinfo{title}{{Diffraction effects in optical interferometric
  displacement detection in nanoelectromechanical systems}}.
\newblock \emph{\bibinfo{journal}{Appl. Phys. Lett.}}
  \textbf{\bibinfo{volume}{86}}, \bibinfo{pages}{013106}
  (\bibinfo{year}{2005}).

\bibitem{ref:Povinelli}
\bibinfo{author}{{Povinelli}, M.~L.} \emph{et~al.}
\newblock \bibinfo{title}{{Evanescent-wave bonding between optical
  waveguides}}.
\newblock \emph{\bibinfo{journal}{Opt. Lett.}} \textbf{\bibinfo{volume}{30}},
  \bibinfo{pages}{3042--3044} (\bibinfo{year}{2005}).

\bibitem{ref:Li_Tang2}
\bibinfo{author}{{Li}, M.}, \bibinfo{author}{{Pernice}, W.~H.~P.} \&
  \bibinfo{author}{{Tang}, H.~X.}
\newblock \bibinfo{title}{{Tunable bipolar optical interactions between guided
  lightwaves}}.
\newblock \emph{\bibinfo{journal}{Nature Photonics}}
  \textbf{\bibinfo{volume}{3}}, \bibinfo{pages}{464--468}
  (\bibinfo{year}{2009}).

\bibitem{ref:Roels}
\bibinfo{author}{{Roels}, J.} \emph{et~al.}
\newblock \bibinfo{title}{{Tunable optical forces between nanophotonic
  waveguides}}.
\newblock \emph{\bibinfo{journal}{Nature Nanotechnology}}
  \textbf{\bibinfo{volume}{4}}, \bibinfo{pages}{510--513}
  (\bibinfo{year}{2009}).

\bibitem{ref:Pruessner}
\bibinfo{author}{{Pruessner}, M.~W.} \emph{et~al.}
\newblock \bibinfo{title}{{End-coupled optical waveguide MEMS devices in the
  indium phosphide material system}}.
\newblock \emph{\bibinfo{journal}{Journal of Micromechanics and
  Microengineering}} \textbf{\bibinfo{volume}{16}}, \bibinfo{pages}{832--842}
  (\bibinfo{year}{2006}).

\bibitem{ref:Li_Tang_cantilever}
\bibinfo{author}{{Li}, M.}, \bibinfo{author}{{Pernice}, W.~H.~P.} \&
  \bibinfo{author}{{Tang}, H.~X.}
\newblock \bibinfo{title}{{Broadband all-photonic transduction of
  nanocantilevers}}.
\newblock \emph{\bibinfo{journal}{Nature Nanotechnology}}
  \textbf{\bibinfo{volume}{4}}, \bibinfo{pages}{377--382}
  (\bibinfo{year}{2009}).

\bibitem{ref:Kippenberg_Vahala_OE}
\bibinfo{author}{{Kippenberg}, T.~J.} \& \bibinfo{author}{{Vahala}, K.~J.}
\newblock \bibinfo{title}{{Cavity Opto-Mechanics}}.
\newblock \emph{\bibinfo{journal}{Opt. Express}} \textbf{\bibinfo{volume}{15}},
  \bibinfo{pages}{17172--17205} (\bibinfo{year}{2007}).

\bibitem{ref:Favero_Karrai}
\bibinfo{author}{{Favero}, I.} \& \bibinfo{author}{{Karrai}, K.}
\newblock \bibinfo{title}{{Optomechanics of deformable optical cavities}}.
\newblock \emph{\bibinfo{journal}{Nature Photonics}}
  \textbf{\bibinfo{volume}{3}}, \bibinfo{pages}{201--205}
  (\bibinfo{year}{2009}).

\bibitem{ref:van_Thorhout}
\bibinfo{author}{{van Thourhout}, D.} \& \bibinfo{author}{{Roels}, J.}
\newblock \bibinfo{title}{{Optomechanical device actuation through the optical
  gradient force}}.
\newblock \emph{\bibinfo{journal}{Nature Photonics}}
  \textbf{\bibinfo{volume}{4}}, \bibinfo{pages}{211--217}
  (\bibinfo{year}{2010}).

\bibitem{ref:Schliesser_resolved_sideband2}
\bibinfo{author}{{Schliesser}, A.}, \bibinfo{author}{{Arcizet}, O.},
  \bibinfo{author}{{Rivi{\`e}re}, R.}, \bibinfo{author}{{Anetsberger}, G.} \&
  \bibinfo{author}{{Kippenberg}, T.~J.}
\newblock \bibinfo{title}{{Resolved-sideband cooling and position measurement
  of a micromechanical oscillator close to the Heisenberg uncertainty limit}}.
\newblock \emph{\bibinfo{journal}{Nature Physics}}
  \textbf{\bibinfo{volume}{5}}, \bibinfo{pages}{509--514}
  (\bibinfo{year}{2009}).

\bibitem{ref:Teufel}
\bibinfo{author}{{Teufel}, J.~D.}, \bibinfo{author}{{Donner}, T.},
  \bibinfo{author}{{Castellanos-Beltran}, M.~A.}, \bibinfo{author}{{Harlow},
  J.~W.} \& \bibinfo{author}{{Lehnert}, K.~W.}
\newblock \bibinfo{title}{{Nanomechanical motion measured with an imprecision
  below that at the standard quantum limit}}.
\newblock \emph{\bibinfo{journal}{Nature Nanotechnology}}
  \textbf{\bibinfo{volume}{4}}, \bibinfo{pages}{820--823}
  (\bibinfo{year}{2009}).

\bibitem{ref:Eichenfield_zipper}
\bibinfo{author}{{Eichenfield}, M.}, \bibinfo{author}{{Camacho}, R.},
  \bibinfo{author}{{Chan}, J.}, \bibinfo{author}{{Vahala}, K.~J.} \&
  \bibinfo{author}{{Painter}, O.}
\newblock \bibinfo{title}{{A picogram- and nanometre-scale photonic-crystal
  optomechanical cavity}}.
\newblock \emph{\bibinfo{journal}{Nature}} \textbf{\bibinfo{volume}{459}},
  \bibinfo{pages}{550--555} (\bibinfo{year}{2009}).

\bibitem{ref:Schliesser_NJP}
\bibinfo{author}{{Schliesser}, A.}, \bibinfo{author}{{Anetsberger}, G.},
  \bibinfo{author}{{Rivi{\`e}re}, R.}, \bibinfo{author}{{Arcizet}, O.} \&
  \bibinfo{author}{{Kippenberg}, T.~J.}
\newblock \bibinfo{title}{{High-sensitivity monitoring of micromechanical
  vibration using optical whispering gallery mode resonators}}.
\newblock \emph{\bibinfo{journal}{New Journal of Physics}}
  \textbf{\bibinfo{volume}{10}}, \bibinfo{pages}{095015}
  (\bibinfo{year}{2008}).

\bibitem{ref:Anetsberger_near_field}
\bibinfo{author}{{Anetsberger}, G.} \emph{et~al.}
\newblock \bibinfo{title}{{Near-field cavity optomechanics with nanomechanical
  oscillators}}.
\newblock \emph{\bibinfo{journal}{Nature Physics}}
  \textbf{\bibinfo{volume}{5}}, \bibinfo{pages}{909--914}
  (\bibinfo{year}{2009}).

\bibitem{ref:Sheard}
\bibinfo{author}{{Sheard}, B.~S.}, \bibinfo{author}{{Gray}, M.~B.},
  \bibinfo{author}{{Mow-Lowry}, C.~M.}, \bibinfo{author}{{McClelland}, D.~E.}
  \& \bibinfo{author}{{Whitcomb}, S.~E.}
\newblock \bibinfo{title}{{Observation and characterization of an optical
  spring}}.
\newblock \emph{\bibinfo{journal}{Phys. Rev. A}} \textbf{\bibinfo{volume}{69}},
  \bibinfo{pages}{051801} (\bibinfo{year}{2004}).

\bibitem{ref:Minne}
\bibinfo{author}{{Minne}, S.~C.} \emph{et~al.}
\newblock \bibinfo{title}{{Centimeter scale atomic force microscope imaging and
  lithography}}.
\newblock \emph{\bibinfo{journal}{Appl. Phys. Lett.}}
  \textbf{\bibinfo{volume}{73}}, \bibinfo{pages}{1742--44}
  (\bibinfo{year}{1998}).

\bibitem{ref:Srinivasan7}
\bibinfo{author}{Srinivasan, K.}, \bibinfo{author}{Barclay, P.~E.},
  \bibinfo{author}{Borselli, M.} \& \bibinfo{author}{Painter, O.}
\newblock \bibinfo{title}{{Optical-fiber-based measurement of an ultrasmall
  volume, high-$Q$ photonic crystal microcavity}}.
\newblock \emph{\bibinfo{journal}{Phys. Rev. B}} \textbf{\bibinfo{volume}{70}},
  \bibinfo{pages}{081306R} (\bibinfo{year}{2004}).

\bibitem{ref:Borselli2}
\bibinfo{author}{Borselli, M.}, \bibinfo{author}{Johnson, T.~J.} \&
  \bibinfo{author}{Painter, O.}
\newblock \bibinfo{title}{{Beyond the Rayleigh scattering limit in high-$Q$
  silicon microdisks: theory and experiment}}.
\newblock \emph{\bibinfo{journal}{Opt. Express}} \textbf{\bibinfo{volume}{13}},
  \bibinfo{pages}{1515--1530} (\bibinfo{year}{2005}).

\bibitem{ref:Verbridge}
\bibinfo{author}{{Verbridge}, S.~S.}, \bibinfo{author}{{Ilic}, R.},
  \bibinfo{author}{{Craighead}, H.~G.} \& \bibinfo{author}{{Parpia}, J.~M.}
\newblock \bibinfo{title}{{Size and frequency dependent gas damping of
  nanomechanical resonators}}.
\newblock \emph{\bibinfo{journal}{Appl. Phys. Lett.}}
  \textbf{\bibinfo{volume}{93}}, \bibinfo{pages}{013101}
  (\bibinfo{year}{2008}).

\bibitem{ref:Caves}
\bibinfo{author}{{Caves}, C.~M.}, \bibinfo{author}{{Thorne}, K.~S.},
  \bibinfo{author}{{Drever}, R.~W.~P.}, \bibinfo{author}{{Sandberg}, V.~D.} \&
  \bibinfo{author}{{Zimmermann}, M.}
\newblock \bibinfo{title}{{On the measurement of a weak classical force coupled
  to a quantum-mechanical oscillator. I. Issues of principle}}.
\newblock \emph{\bibinfo{journal}{Rev. Mod. Phys.}}
  \textbf{\bibinfo{volume}{52}}, \bibinfo{pages}{341--392}
  (\bibinfo{year}{1980}).

\bibitem{ref:Corbitt}
\bibinfo{author}{{Corbitt}, T.} \emph{et~al.}
\newblock \bibinfo{title}{{Optical Dilution and Feedback Cooling of a
  Gram-Scale Oscillator to 6.9mK}}.
\newblock \emph{\bibinfo{journal}{Phys. Rev. Lett.}}
  \textbf{\bibinfo{volume}{99}}, \bibinfo{pages}{160801}
  (\bibinfo{year}{2007}).

\bibitem{ref:Johnson_TJ}
\bibinfo{author}{{Johnson}, T.~J.}, \bibinfo{author}{{Borselli}, M.} \&
  \bibinfo{author}{{Painter}, O.}
\newblock \bibinfo{title}{{Self-induced optical modulation of the transmission
  through a high-Q silicon microdisk resonator}}.
\newblock \emph{\bibinfo{journal}{Opt. Express}} \textbf{\bibinfo{volume}{14}},
  \bibinfo{pages}{817--831} (\bibinfo{year}{2006}).

\bibitem{ref:Albrecht_FM_AFM}
\bibinfo{author}{{Albrecht}, T.~R.}, \bibinfo{author}{{Gr{\"u}tter}, P.},
  \bibinfo{author}{{Horne}, D.} \& \bibinfo{author}{{Rugar}, D.}
\newblock \bibinfo{title}{{Frequency modulation detection using high-Q
  cantilevers for enhanced force microscope sensitivity}}.
\newblock \emph{\bibinfo{journal}{J. Appl. Phys.}}
  \textbf{\bibinfo{volume}{69}}, \bibinfo{pages}{668--673}
  (\bibinfo{year}{1991}).

\bibitem{ref:Srinivasan12}
\bibinfo{author}{Srinivasan, K.}, \bibinfo{author}{Borselli, M.},
  \bibinfo{author}{Painter, O.}, \bibinfo{author}{Stintz, A.} \&
  \bibinfo{author}{Krishna, S.}
\newblock \bibinfo{title}{{Cavity $Q$, mode volume, and lasing threshold in
  small diameter AlGaAs microdisks with embedded quantum dots}}.
\newblock \emph{\bibinfo{journal}{Opt. Express}} \textbf{\bibinfo{volume}{14}},
  \bibinfo{pages}{1094--1105} (\bibinfo{year}{2006}).

\bibitem{ref:Stowe_aN_force}
\bibinfo{author}{{Stowe}, T.~D.} \emph{et~al.}
\newblock \bibinfo{title}{{Attonewton force detection using ultrathin silicon
  cantilevers}}.
\newblock \emph{\bibinfo{journal}{Appl. Phys. Lett.}}
  \textbf{\bibinfo{volume}{71}}, \bibinfo{pages}{288--290}
  (\bibinfo{year}{1997}).

\bibitem{ref:Rugar_single_spin}
\bibinfo{author}{{Rugar}, D.}, \bibinfo{author}{{Budakian}, R.},
  \bibinfo{author}{{Mamin}, H.~J.} \& \bibinfo{author}{{Chui}, B.~W.}
\newblock \bibinfo{title}{{Single spin detection by magnetic resonance force
  microscopy}}.
\newblock \emph{\bibinfo{journal}{Nature}} \textbf{\bibinfo{volume}{430}},
  \bibinfo{pages}{329--332} (\bibinfo{year}{2004}).

\bibitem{ref:Torbrugge}
\bibinfo{author}{Torbr{\"{u}}gge, S.}, \bibinfo{author}{Schaff, O.} \&
  \bibinfo{author}{Rychen, J.}
\newblock \bibinfo{title}{{Application of the KolibriSensor$^{\textregistered}$
  to combined atomic-resolution scanning tunneling microscopy and noncontact
  atomic-force microscopy}}.
\newblock \emph{\bibinfo{journal}{J. Vac. S. Tech. B}}
  \textbf{\bibinfo{volume}{28}}, \bibinfo{pages}{C4E12--C4E20}
  (\bibinfo{year}{2010}).

\bibitem{ref:Giessibl_QPLus}
\bibinfo{author}{{Giessibl}, F.~J.}
\newblock \bibinfo{title}{{High-speed force sensor for force microscopy and
  profilometry utilizing a quartz tuning fork}}.
\newblock \emph{\bibinfo{journal}{Appl. Phys. Lett.}}
  \textbf{\bibinfo{volume}{73}}, \bibinfo{pages}{3956--3958}
  (\bibinfo{year}{1998}).

\bibitem{ref:Wu_Solgaard_Ford}
\bibinfo{author}{{Wu}, M.~C.}, \bibinfo{author}{{Solgaard}, O.} \&
  \bibinfo{author}{{Ford}, J.~E.}
\newblock \bibinfo{title}{{Optical MEMS for Lightwave Communication}}.
\newblock \emph{\bibinfo{journal}{J. Lightwave Tech.}}
  \textbf{\bibinfo{volume}{24}}, \bibinfo{pages}{4433--4454}
  (\bibinfo{year}{2006}).

\bibitem{ref:Senturia}
\bibinfo{author}{Senturia, S.~D.}
\newblock \emph{\bibinfo{title}{{Microsystem Design}}}
  (\bibinfo{publisher}{Springer Science+Business Media}, \bibinfo{address}{New
  York, NY}, \bibinfo{year}{2004}), \bibinfo{edition}{7th printing} edn.

\end{thebibliography}
\end{document}